\documentclass[copyright,creativecommons]{eptcs}

\usepackage{times}
\usepackage{graphicx}
\usepackage{listings}
\usepackage{alltt}
\usepackage{amsmath,amssymb,amsthm}
\usepackage{array}
\usepackage[T1]{fontenc}
\usepackage{verbatim}
\usepackage{color}
\usepackage{url}
\usepackage{multirow}
\usepackage{breakurl}   
\usepackage{eucal}
\usepackage{xspace}

\graphicspath{{images/}}

\theoremstyle{definition}
\newtheorem{definition}{Definition}

\renewcommand{\paragraph}[1]{\medskip \noindent {\bf #1}}

\newcommand{\id}[1]{\mathit{#1}}

\newcommand{\M}{\mathcal{M}}

\newcommand{\Act}{\textit{Act}\xspace}

\newcommand{\Uppaal}{{\sc Uppaal}\xspace}

\newcommand{\scaleIndex}{.9}
\newcommand{\trule}[1]{\big[\hspace{-.1cm}\underline{\big[#1\big]}\hspace{-.1cm}\big]}
\newcommand{\tneg}{\sim\!}

\lstdefinelanguage{Credo}
{morekeywords={interface, class, facade, component, manager, inherits, implements, begin, end, while, do, od, if, fi, then, else, and,  op, with, in, out, inout, or, var, port, new, inports, return, int, bool, outports, syncports, sync_event, async_event, await, release, raise_event, skip},
sensitive=true,
morecomment=[l]{//},
morecomment=[s]{/*}{*/},
morestring=[b]",
}
\lstdefinelanguage{maude}
{morekeywords={mod, is, pr, op, eq, rew, endm},
sensitive=true,
morecomment=[l]{***},
morecomment=[s]{/*}{*/},
morestring=[b]",
}
\lstdefinelanguage{Uppaal}
{morekeywords={while, do, if, else, and,  or, new, return, int, bool, clock, void, const, meta},
sensitive=true,
morecomment=[l]{//},
morecomment=[s]{/*}{*/},
morestring=[b]",
}

\begin{document}

\title{Timed Automata Semantics for Analyzing Creol\thanks{This work is partly funded by the European IST-33826 STREP project CREDO on Modeling and Analysis of Evolutionary Structures for Distributed Services.}}

\def\titlerunning{Timed Automata Semantics for Analyzing Creol}  

\author{
Mohammad Mahdi Jaghoori
\institute{LIACS, Leiden University, Leiden\\
CWI, Amsterdam, The Netherlands}
\email{jaghouri@cwi.nl}
\and
Tom Chothia
\institute{School of Computer Science\\
University of Birmingham, United Kingdom}
\email{t.chothia@cs.bham.ac.uk}
}

\def\authorrunning{M. Jaghoori and T. Chothia}   

\maketitle

\begin{abstract}
We give a real-time semantics for the concurrent, object-oriented modeling language Creol, by mapping Creol processes to a network of timed automata. We can use our semantics to verify real time properties of Creol objects, in particular to see whether processes can be scheduled correctly and meet their end-to-end deadlines.
Real-time Creol can be useful for analyzing, for instance, abstract models of multi-core embedded systems.
We show how
analysis can be done
in \Uppaal.

\end{abstract}

\section{Introduction}
Parallel and distributed systems span a wide range of applications, from internet-based services to multi-core embedded systems.
Concurrent objects, as in Creol \cite{johnsen07sosym}, have dedicated processors and thus provide a natural way of modeling these systems.
In a real-time setting, a major analysis problem is schedulability, that is, whether all tasks are accomplished within their deadlines.
We enrich Creol with real-time and provide  a timed-automata semantics for it where in particular schedulability can be analyzed; this semantics allows for explicit modeling of  scheduling strategies.
For practicality, we define this semantics in terms of a translation algorithm such that the generated automata can be handled by \Uppaal\ \cite{LarsenPY97}.




Each Creol method is mapped to exactly one automaton with the possibility of tracing from automata back to Creol code; the semantics of an object consists of a network of the automata for its methods together with a scheduler automaton (e.g., Earliest Deadline First, Fixed Priority Scheduling, etc.).
We have implemented a program that translates Creol code to the corresponding timed automata, which can then be run and verified in \Uppaal. We validate the automated verification cycle by checking a coordinator class by Johnsen and Owe \cite {johnsen07sosym}, augmented with real-time, for correctness as well as schedulability.


A distinctive feature of Creol is that it gives the programmer a high degree of control over the scheduling. At ``processor release points'', a method can choose to give up control of the processor and specify conditions on when it should be resumed.
Previous work on Creol has not addressed particular scheduling policies; instead processes are scheduled in a completely non-deterministic manner. The real time extensions and schedulability analysis we present here can be seen as a complementary feature for the basic Creol language.

In Creol, object references are typed only by interfaces, allowing for a strongly typed language with support for features such as multiple inheritance and type-safe dynamic class upgrades \cite{yu06fmoods}. We augment an interface with a notion of high-level behavioral specification given in a timed automaton; this is given by the modeler as the first step in modeling an object. A class can implement multiple interfaces. In this case, the timed automata for behavioral interfaces should be interleaved; intuitively, allowing more behavior than each automaton separately. The justification is that a class with multiple interfaces can participate in multiple protocols independently. Creol supports multiple inheritance for both interfaces and classes.  Interfaces form a subtype hierarchy, which is distinct from inheritance at the level of classes used merely for code reuse \cite{JohnsenOY06}. In a subtype hierarchy, the timed automata for the inherited behavioral interfaces should synchronize on similar actions; intuitively, allowing less behavior in the subtype. The justification is that a subtype should be a refinement of its supertypes \cite{PR94}.



We use a statement-based delay semantics for our notion of real-time in Creol, namely, we annotate each statement with best- and worst-case execution times, as it naturally fits the operational semantics of Creol and it is the normal trend in schedulability analysis \cite{EDF08,ClossePPSVWY01,FersmanYi07acc,KloukinasY03}.
Also, all method calls are given a deadline, which specifies the time before which the callee {\it should} finish.
By giving the annotations in comments, we allow the standard Creol interpreter to behave normally by ignoring the real-time information if necessary.
One could specify cumulative delays for a group of statements (by assigning the default delay of zero to all statements except the last one).
When a Creol model is schedulable, it gives us requirements on point-to-point execution times that should be fulfilled by the final implementation in programming language like C; this can also be tested by checking conformance \cite{AichernigGJSS08}. 
Whenever a called method finishes, it sends a `reply' back to the caller. By waiting for the reply, one can model synchronous method calls. Timely replies ensure end-to-end deadlines. Method invocations are labeled to help match replies with the originating invocations.
Dynamic labels lead to infinite state models for non-terminating systems. To model this using finite state timed automata, we label method invocations statically;
thus, replies to repeated invocations associated with the same label are not distinguished in our automata semantics.
Furthermore, we abstract from other dynamic statements in Creol.
Although in our semantics, we abstract from the dynamic behavior, i.e., dynamic object creation and reconfiguration, it is worth noting that these actions have no effect in individual object analysis.

In Section \ref{sec:prelim}, we briefly introduce timed automata and \Uppaal models.
Section \ref{sec::rtcreol} describes Creol and discusses how it can be extended with real time.
In Section \ref{sec::translation} we discuss how Creol with real time can be given a semantics using timed automata.
We show how this semantics can be applied to analyze schedulability in Section \ref{sec::analysis}.
We conclude in Section \ref{sec:conc}.

\paragraph{\bf Related Work}
In previous work \cite{Jaghoori09jlap,JaghouriDBC08RTSS}, we provided a high-level framework for modular schedulability analysis of purely asynchronous objects modeled as timed automata. This past work assumes methods run until completion without interruption, unlike the work we present here. In this paper we apply this automata theoretic framework to the concurrent object modeling language Creol \cite{johnsen07sosym} by defining a mapping from Creol processes to timed-automata; to accomodate Creol features, the scheduler model is also extensively extended.

We discussed a possible encoding from Creol to timed automata in a past paper \cite{fsen09}, however this encoding was incomplete, both in the Creol features covered and because it discussed only translation of one method; it would generate  many automata to encode each method.
In the full, efficient encoding presented in this paper, we improve scalability by reducing the number of generated automata (i.e., release statements do not generate a new automaton now). Further, we improve label handling by making the scheduler responsible for this; we add features like multiple interfaces; blocking statements are now also handled by the scheduler; our improved scheduler allows for local synchronous calls, which may incur recursive calls. With this semantics, we can now verify Creol models for schedulability; examples of other properties we can check include deadlock, or timed reachability of any specific line of code.

${\rm Creol_{RT}}$ \cite{KyasJ08} is a real-time extension of Creol along the lines of   Hoare logic extended with real-time.
In ${\rm Creol_{RT}}$, time can affect the functional behavior of the object, but we use only descriptive annotations.
It is possible in ${\rm Creol_{RT}}$ to specify contradicting invariants, which is not the case in our simple delay model.
By automatically generating timed automata, we can readily use \Uppaal\ for model checking and schedulability analysis, however, to the best of our knowledge, ${\rm Creol_{RT}}$ has no automated tool support.
Another important difference is that we can model and compare explicit scheduling strategies, and so say which strategies would and would not lead to missed deadlines.

With respect to schedulability analysis, a characteristic of our work is modularity. A behavioral interface models the most general input/output behavior allowed for an object and thus can be used as an abstraction of the environment. A behavioral interface can be viewed as a contract as in ``design by contract'' \cite{Meyer92} or as a most general assumption in modular model checking ~\cite{KupfermanVW01} (based on assume-guarantee reasoning); schedulability is guaranteed if the real use of the object satisfies this assumption.
In the literature,
a model of the environment is usually the task generation scheme in a specific situation. For example, in TAXYS \cite{ClossePPSVWY01}, different models of the environment can be used to check schedulability of the application in different situations. However, a behavioral interface in our analysis
covers all allowable usages of the object, and is thus an over-approximation of all environments in which the object can be used.
This adds to the modularity of our approach; every use of the object foreseen in the interface is verified to be schedulable.
Finally, TAXYS deals with a programming language, whereas we work at modeling level. It is also the case with the works of \cite{KloukinasY03} where they extract automata from code for schedulability analysis. As mentioned above, they deal with programming languages (like TAXYS) and timings are usually obtained by profiling the real system. Our work, on the contrary, is applied on a model before the implementation of the real system. Therefore, our main focus is on studying different scheduling policies and design decisions.
These models can be used for automatic code generation or conformance checking with later or existing implementations.

\section{Preliminaries}
\label{sec:prelim}

We give a brief introduction to Timed Automata, for a full description we refer the reader to previous papers on timed automata (e.g. \cite{AlurD94}) and documentation for the \Uppaal tool \cite{LarsenPY97}.

\begin{definition}[\bf Timed Automata]\label{def:timeAut}
Suppose $\mathcal{B}(C)$ is the set of all clock constraints on the set of clocks $C$. A timed automaton over actions $\Act$ and clocks $C$ is a tuple $\langle L, l_0, \longrightarrow, I \rangle$ representing
\begin{itemize}
\item a finite set of locations $L$ (including an initial location $l_0$);
\item the set of edges $\longrightarrow \subseteq L \times \mathcal{B}(C) \times \Act \times 2^C \times L$; and,
\item a function $I: L \mapsto \mathcal{B}(C)$ assigning an invariant to each location.
\end{itemize}
\end{definition}

An edge  $(l ,g,a,r, l')$ implies that action `$a$' may change the location $l$ to $l'$ by resetting the clocks in $r$,  if the clock constraints in $g$ (as well as the invariant of $l'$) hold.
A timed automaton is called {\em deterministic} if and only if for
each $a \in \Act$, if there are two edges $(l ,g,a,r, l')$ and $(l ,g',a,r', l'')$ from $l$ labeled by the same
action $a$
then the guards $g$ and $g'$ are disjoint (i.e., $g \wedge g'$ is unsatisfiable).
Since we use \Uppaal~\cite{LarsenPY97}, we allow defining variables of type boolean
and bounded integers. Variables can appear in
guards and updates.




A system may be described as a {\em network} of communicating timed automata. 
In these automata, the action set is partitioned into input, output and internal actions.
The behavior of the system is defined as the parallel composition of those automata $A_1 \parallel \dots \parallel A_n$. Semantically, the system can delay if all automata can delay and can perform an action if one of the automata can perform an internal action or if two automata can synchronize on complementary actions (inputs and outputs are complementary).
In a network of timed automata, variables can be defined locally for one automaton, globally (shared between all automata), or as  parameters to the automata.

A location can be marked  {\em urgent} in an automaton to indicate that the automaton cannot spend any time in that location. This is equivalent to resetting a fresh clock $x$ in all of its incoming edges and adding an invariant $x \le 0$ to the location.
In a network of timed automata, the enabled transitions from an urgent location may be interleaved with the enabled transitions from other automata (while time is frozen).
Like urgent locations, {\em committed} locations freeze time; furthermore, if any process is in a committed location, the next step must involve an edge from one of the committed locations.

\begin{definition}[\bf {\textsc {\bfseries Uppaal}} Model]
An \Uppaal\ model consists of: (1)  a set of timed automata templates ($\id{TAT}$); (2) global declarations; and, (3) system declarations.
\end{definition}

An automata template in an \Uppaal model consists of a name, a set of arguments, local declarations and a timed automaton definition (as above); formally, $\id{TAT} = (\id{tName}, \id{Args}, \id{local}, \id{Auto})$.
Global and local declarations contain the definition of clocks and variables.
The network of timed automata to be analyzed is defined in the system declarations by instantiating the timed automata templates.

\section{Creol with Real-Time}
\label{sec::rtcreol}

\lstset{language=Credo, basicstyle=\scriptsize\ttfamily} 

\begin{figure}[tb]
\begin{tabular}{ll}
{ \small 
\begin{tabular}{l@{~}c@{~~}l}
b &:& Boolean \\
e &:& Expression \\
g&:& Guard  \\
m&:&Method \\
n &:& Identifier \\
s &:& Statement \\
t &:& Label \\
x &:& Object \\

\end{tabular}
}
&
{\small
\begin{math}
\begin{array}{l@{~~}r@{~~}l}
\id{in}\ &::=& \id{\bf interface} ~ n([n:n]^*_,) ~[\id{\bf inherits} ~[n[(e^+_,)]^?]^+_,]^?
~ \id{\bf begin} ~ [[\id{\bf with}~ n]^?~\id{\bf op} ~n]^+ ~\id{\bf end}\\
cl\ &::=& \id{\bf class} ~ n([n:n]^*_,) ~ \id{\bf implements}~ [n[(e^+_,)]^?]^+_,
~ \id{\bf begin} ~ [\id{\bf var}\  \id{Vdcl}]^* ~[mtd]^+ ~\id{\bf end}\\
\id{Vdcl}& ::= & n^+_, : [int ~|~ bool] \\
mtd & ::=& [\id{\bf with}~ n]^?~\id{\bf op} ~n~ ==~ S \\
S\ & ::= & s \ | \ s;S \\
s\ &::= & n := e\ |\ !x.m()\ | \ t!x.m() \ | \ t? ~|\ \id{\bf release}\ |\  \id{\bf await}\ g ~\\
&&| \  \id{\bf while}\ b\ \id{\bf do}\ S\  \id{\bf od}
~ | \ \id{\bf if}\ b\ \id{\bf then}\ S\ \id{\bf else}\ S ~|~ \id{\bf skip} \\ 
g\ &::=& b\ |\ t?\ |\ \tneg g\ |\ g \land g \\
\end{array}
\end{math}
}
\end{tabular}
\caption{BNF  grammar for Creol (adapted from \cite{johnsen07sosym}) where $^*$ and $^+$ show repetition for at least 0 and 1 times, respectively; $^?$ denotes an optional element; and, a subscript $_,$ implies a comma-separated list.}\label{fig::creolSyntax}
\end{figure}

\begin{figure}[t]

\begin{tabular}{m{.60\textwidth}m{.30\textwidth}}

\begin{lstlisting}
interface C1 begin with Any op m1 end
interface C2 begin with Any op m2 end
interface C3 begin with Any op m3 end
class Coordinator implements C1, C2, C3 begin
  var s1,s2,s3,sync : bool
  op init ==
    s1 := false; // assume zero exec. time by
    s2 := false; // default, resulting in a
    s3 := false; // collective delay
    sync := true /* specified here: @b2 @w4*/
  op body ==
    skip                          /*@b2 @w5*/
  op run ==
    await (s1 /\ s2 /\ s3);       /*@b1 @w2*/
    b!body();  /* invoc. delay @b1 @w1 @d10*/
    b?;        // force sync call
    sync := false;
    await (~s1 /\ ~s2 /\ ~s3);    /*@b1 @w2*/
    sync := true;                 /*@b1 @w1*/
    !run()                        /*   @d50*/
  with Any
  op m1 ==
    await (sync /\ ~s1);
    s1 := true;                   /*@b1 @w1*/
    await ~sync;
    s1 := false                   /*@b1 @w1*/
  op m2 ==  ...like m1...
  op m3 ==  ...like m1...
end
\end{lstlisting}

&

{\small
\centering
\includegraphics{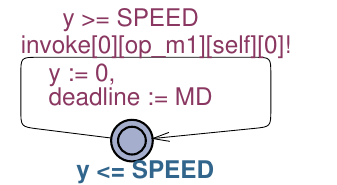} \\ Behavioral specification for C1 \\ ~ \\~ \\
\includegraphics{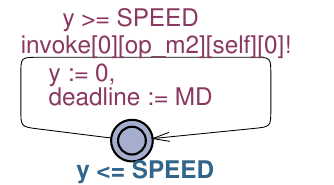} \\ Behavioral specification for C2 \\ ~ \\~ \\
\includegraphics{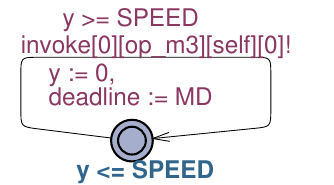} \\ Behavioral specification for C3
}

\end{tabular}
\vspace{-0.5cm}
\caption{A real-time coordinator in Creol with periodic task generation in the behavioral interfaces.}
\label{fig::creolCode}\label{fig::periodicBeh}

\end{figure}

Creol~\cite{johnsen07sosym} is an object oriented modeling language for distributed systems, where each object implicitly has a dedicated processor.
A simplified syntax for Creol, for which we give a semantics is shown in Figure~\ref{fig::creolSyntax}.
We use, as running example, a coordinator class, taken from \cite{johnsen07sosym}, concretized for three-way synchronization (while abstracting data away), shown in Figure~\ref{fig::creolCode}.

A Creol model is defined as objects typed by interfaces, which are in turn a set of method definitions; a method can define a cointerface using the \lstinline{with} keyword to restrict the type of its caller to the given interface.
A class implementing some interfaces realizes their methods possibly introducing private methods and local variables.
With inheritance of interfaces, one can create a subtype hierarchy.
We abstract from method parameters and dynamic object creation.
However, classes and interfaces can have parameters. Instances of a class can communicate by objects given as class parameters, called the {\em known objects}. We  can thus define the static topology of the system.
The class behavior is defined in its methods, where a method is a sequence of statements separated by semicolon.
For expressions, we assume the syntax that is accepted by \Uppaal.
The \lstinline{Coordinator} class implements three interfaces, each with one method, e.g., \lstinline{with Any op m1} that defines method \lstinline{m1} which can be called from objects of {\em any} type.
The method \lstinline{init} in an object is immediately executed upon object creation.
The method \lstinline{run} specifies active behavior of the object; it will have to compete with other tasks in the queue if any.

\lstset{language=Uppaal, basicstyle=\footnotesize}

Methods can have processor release points which define interleaving points explicitly.
When a process is executing, it is not interrupted until it finishes or reaches a release point.
Release points can be conditional, written as \lstinline $await g$ (e.g., line 14).
If \lstinline $g$ is satisfied, the process keeps the control; otherwise, it releases  the processor.
A suspended process will be enabled when the guard evaluates to true.
When the processor is free, an enabled process is nondeterministically selected and started.
The \lstinline $release$ statement unconditionally releases the processor and the continuation of the  method is an enabled process.

If a method invocation \lstinline $p$ is associated with a label \lstinline $t$, written as \lstinline $t!p()$, the sender can wait for a reply using the blocking statement \lstinline $t?$ or in a nonblocking way by including \lstinline $t?$ in a release point, e.g., as in  \lstinline$await t?$.
A reply is sent back automatically when the called method finishes.
Before the reply is available, executing  \lstinline$await t?$ releases the processor whereas  the blocking statement \lstinline $t?$ does not. 
While the processor is not released, the other processes in the object do not get a chance for execution; if \lstinline$t?$ is related to a self call and its reply is not yet available, it forces synchronous execution of the called method.
For example, line 16 in Figure \ref{fig::creolCode} forces synchronous execution of the method \lstinline{body}.

\paragraph{Adding Real-Time}
The modeler should specify for every statement how long it takes to execute.
The directives \lstinline{@b} and \lstinline{@w} are used for specifying the best-case and worst-case execution times for each statement.
We assume zero execution time for statements with no annotations; there must be, however, non-zero execution time in each process before the processor is released.
Furthermore, every method call, including self calls, must be associated with a deadline using \lstinline{@d} directive. This deadline specifies the relative time before which the corresponding method should be scheduled and executed. Since we do not have message transmission delays, the deadline expresses the time until a reply is received. Thus, it corresponds to an {\em end-to-end} deadline.
A worst-case execution time delay for a blocking statement \lstinline$t?$ is ignored.
The delay associated to release statements specifies only the time for invoking the command and not the waiting time afterwards.



Creol interfaces are enriched with behavioral interfaces given in timed automata, which specify
the abstract behavior of an object. 
This interface consists of the messages the object may receive and send and provides an overview of the object behavior in a single automaton. It should also contain the reply signals the object may receive.
A behavioral interface abstracts from specific method implementations, the queue in the object and the scheduling strategy.
Figure \ref {fig::periodicBeh} shows the behavioral interfaces for the coordinator example that  assume  periodic arrival of events.
To formally define behavioral interfaces in \Uppaal, we assume two finite global sets $\M$ for method names and $\mathcal{T}$ for labels; and, two \Uppaal channels \lstinline{invoke} and \lstinline{reply} (cf. modeling schedulers in the next section).

\begin{definition}[\bf Behavioral interface]\label{def:behavioral_interface}
A behavioral interface $B$, with known objects $\mathcal{K}$, providing a set of method names $M_B \subseteq \M$ is formally defined as a deterministic timed automaton over alphabet $\Act^B$  such that $\Act^B$ is partitioned into three sets of actions (assume $t \in \mathcal{T}$ and $k \in \mathcal{K}$):
\begin{itemize}
\item object inputs sent by the environment:
~ $ \Act_I^B = \{\lstinline+invoke[0][+m\lstinline+][self][+k\lstinline+]!+|m \in M_B \}$
\item object outputs received by the environment:
~ $\Act_O^B = \{\lstinline+invoke[+t\lstinline+][+m\lstinline+][+k\lstinline+][self]?+|m \in \M \wedge m \not \in M_B \}$
\item replies to the object outputs:
~ $\Act_r^B = \{\lstinline+reply[+t\lstinline+][self]!+\}$
\end{itemize}
Transitions specifying an input to the object must update the variable \lstinline$deadline$ with an integer specifying the deadline for that input.
If $B$ has no known objects then $\mathcal{K} = \{0\}$ by default.
\end{definition}

\section{Timed Automata Semantics For Creol}\label{sec::translation}

The semantics of a Creol class consists of the automata for its methods. 
When instantiating a class, it should also be associated with a scheduler automaton.
In this section, we explain the algorithm for automatically deriving automata from Creol code.
This semantics maps each method to one automaton.
For practicality, we define this semantics in terms of a translation algorithm
that generates an \Uppaal model containing the automata for methods together with other necessary global declarations.

\begin{definition}[\bf Class]\label{def:class}
The semantics of a class $R$ implementing a set of behavioral interfaces $B^*$ is defined as follows.
Recall that each $B \in B^*$ has a set of method names $M_B$.
$R$ is a set $\{(m_1,E_1,A_1), \ldots, (m_n,E_n,A_n)\}$ of tasks, where
\begin{itemize}
\item $M_R=\{m_1,\dots,m_n\} \subseteq \M$ is a set of task names such that $ M_B \subseteq M_R$ for every $B \in B^*$.
$M_R$ includes the sub-tasks created at release points, as well as the methods; however, there is exactly one automaton for each method.
\item for all $i$, $1 \leq i \leq n$, $A_i$ is a timed automaton representing the method containing the task $m_i$. 
\item for all $i$, $1 \leq i \leq n$, $E_i$ is the enabling condition for $m_i$.
\end{itemize}
\end{definition}

\begin{figure*}[tb]
\begin{center}
\begin{tabular}{@{} m{11.6cm}  m{4.35cm}}
\includegraphics[scale=0.9]{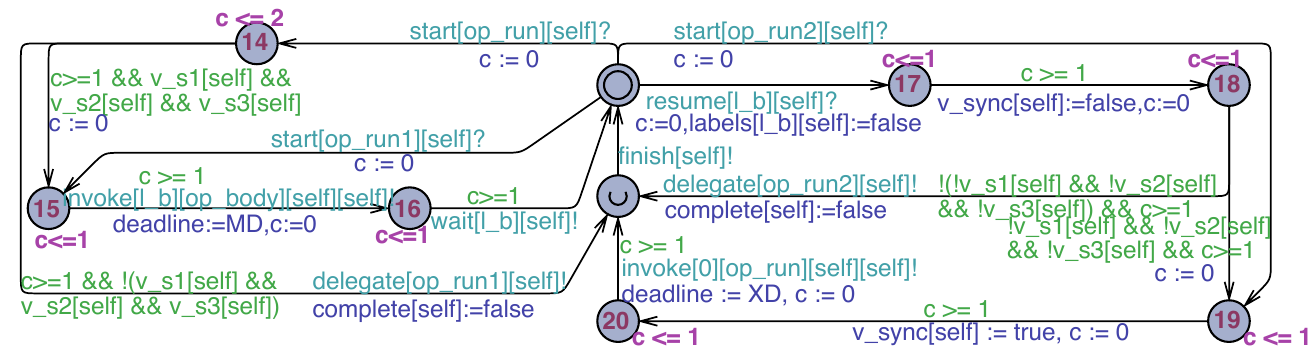} &
{\footnotesize\bfseries
\renewcommand{\arraystretch}{1.25}
\begin{tabular}{@{}l@{~}|@{~}c@{}}
Task & Enabling Condition\\
\hline
\verb"run" & ${\rm true}$\\
\verb"run1" & ${\rm s1 \land s2 \land s3}$\\
\verb"run2" & ${\rm \tneg s1~ \land \tneg s2 ~\land \tneg s3}$ \\
\end{tabular}
\vspace{.5cm}
}
\end{tabular}
\\
(a) The automaton for method {\tt run} and the enabling conditions for its subtasks. \\
\end{center}


%
%
%
\centering
\begin{tabular}{@{} m{8.2cm} >{\centering\arraybackslash} m{4.35cm}}
\includegraphics[scale=0.9]{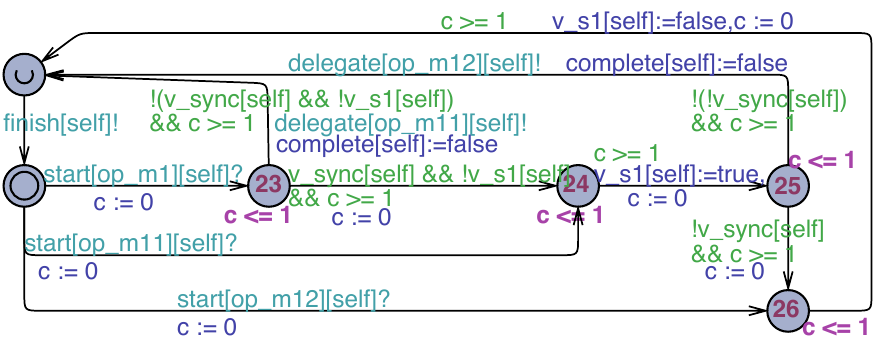}
 &
{\footnotesize\bfseries
\renewcommand{\arraystretch}{1.25}
\begin{tabular}{l|c}
Task & Enabling Condition\\
\hline
\verb"m1" & ${\rm true}$\\
\verb"m11" & ${\rm sync~ \land \tneg s1}$\\
\verb"m12" & ${\rm \tneg sync}$\\
\end{tabular}
}
\end{tabular}
\\
(b) The automaton for method {\tt m1} and the enabling conditions for its subtasks. 
\caption{The labels on automata location refer to line numbers in Figure~\ref{fig::creolCode} (cf. function $\id{Loc}$ in Table \ref{tab:function}).}\label{fig::run}
\end{figure*}

We assume that the given Creol models are correctly typed and annotated with timing information.
We use the same syntax for expressions and assignments in Creol, as is used by \Uppaal. This allows for a more direct translation.
For the sake of simplicity, we abstract from parameter passing, however, it can be modeled in \Uppaal\ by extending the queue to hold the parameters (cf. Section \ref{sec::scheduler}).
Figure~\ref{fig::run}  shows the automata generated for the methods \lstinline$run$ and \lstinline$m1$ from Figure~\ref{fig::creolCode}.
The automata for \lstinline$m2$ and \lstinline$m3$ are similar.
As can be seen in these automata, the method and variable names are prefixed in order to avoid name clashes with \Uppaal\ keywords.
As explained in Section \ref{sec:formal}, function $\id{Loc}$ will link each automata location to its corresponding part of the original code;
this is shown in Figure~\ref{fig::run} as line numbers refering to the original Creol code in Figure~\ref{fig::creolCode}.

Methods may  release the processor before their completion, as in the `\lstinline{await}' statement in line 14 in Figure \ref{fig::creolCode}.
In these cases, the rest of the method is modeled as a sub-task, e.g., the  transition from location 14 to \verb"U" in Figure \ref{fig::run} generates a sub-task \lstinline{op_run1} if the guard associated to \lstinline{await} does not hold; the processor is released by the subsequent transition with the action \lstinline{finish} going back to the initial location.
This sub-task inherits the remaining deadline of the original task; this is done by the scheduler when handling the \lstinline{delegate} channel (see next subsection).
The enabling condition of the sub-tasks are equivalent to the guard used in the corresponding release point (see tables in Figure \ref{fig::run}).
The sub-task \lstinline{op_run1} can be triggered with the \lstinline{start} transition entering location 15.

In standard Creol, different invocations of a method call are associated with different values of the label. For instance executing the statement \lstinline$t!p()$ twice results in two  instances of the label \lstinline $t$. Dynamic labels give rise to an infinite state space for  non-terminating reactive systems.
To be able to use model checking, we treat every label as a static tag. Therefore, different invocations of a method call with the same label are not distinguished in our framework.
Alternatively, one could associate replies to message names, but this is too restrictive.
By associating replies to labels, we can still distinguish the same message sent from different methods with different labels.

We handle labels with a global boolean array \lstinline{labels}, and assuming that label names are unique in each class, we can define constants for each label such that \lstinline{labels[t][self]} uniquely identifies label $t$.
When the condition in a release point includes \lstinline$t?$, i.e., waiting until the reply to the call with label $t$ is available, we replace $t$ with \lstinline{labels[t][self]} which is set to true by the scheduler when called method finishes.
For outgoing messages, the behavioral interface should capture when a reply is expected (cf. Definition \ref{def:behavioral_interface}).

Blocking statements use the \lstinline{wait} channel to transfer the control to the scheduler, which then can check whether the label is associated to a local call; thus it can decide whether to block the object or do a synchronous call.
To allow recursive synchronous calls, the \lstinline{wait} transition goes back to the initial state making the method reentrant, e.g., the transition from location 16 modeling the statement \lstinline{b?}.

Another complication in translation is how to map a possibly infinite state Creol model to finite state automata.
We do this by abstracting away some information. One automatic way of abstracting is as follows:  variables from a finite domain can be mapped to themselves but conditions on potentially infinite variables are mapped to true, we perform this with the function  ${\it Abs}$ in Figure \ref{fig::generation}. We do it semi-automatically, i.e., the user states which variables are abstracted away. In other words, we over-approximate the behavior of the Creol model. A more advanced abstraction would map potentially infinite variables to finite domains in order to narrow the over-approximation.

\subsection{Modeling schedulers}
\label{sec::scheduler}

\begin{figure}
\centering
 \includegraphics[scale=.9]{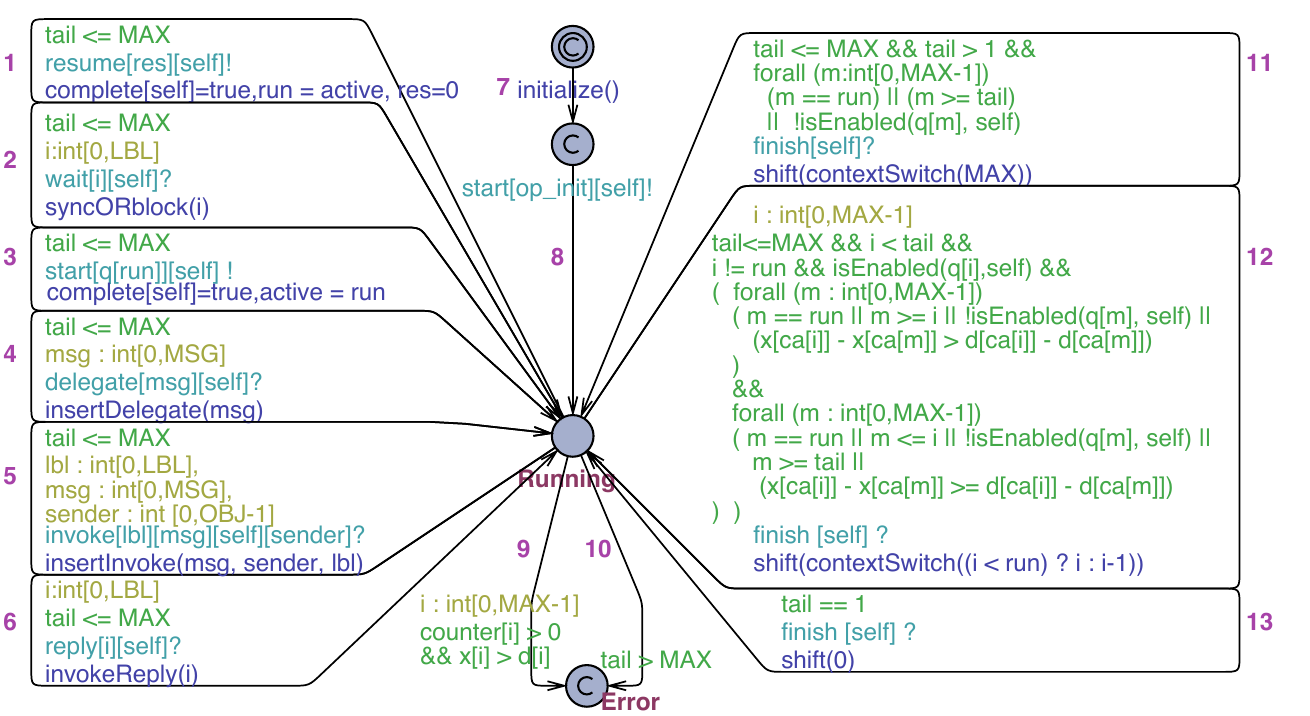}
 \caption{An Earliest-Deadline-First scheduler modeled in \Uppaal (details explained in the text).}
 \label{fig:scheduler}
\end{figure}

A scheduler manages the buffering and execution of the tasks.
A typical {\em scheduler automaton} is given in Figure \ref{fig:scheduler}.
A transition numbered $\alpha$ in this figure is referred to as $t_\alpha$ in the text.
This model is generic except for the strategy, explained below.
The keyword \lstinline{self} holds the current object identifier.
Local declarations including function definitions are not given due to space limitations.

\paragraph{Queue}
The queue has the size \lstinline$MAX$.
A scheduler automaton uses multiple arrays of size \lstinline$MAX$ to implement the queue and its clocks. The message in \lstinline$q[i]$ is received from object \lstinline$s[i]$; and, \lstinline$ca[i]$ points to a clock in array \lstinline$x$ that records how long the message has been in the queue: this message should be processed before the clock \lstinline$x[ca[i]]$ reaches its deadline value \lstinline$d[ca[i]]$.
A clock \lstinline$x[i]$ may be assigned to more than one task, the count of which is stored in \lstinline$counter[i]$.
We have shown in \cite{Jaghoori09jlap} that schedulable objects need bounded queues, i.e., with a proper \lstinline$MAX$ value, queue overflow implies nonschedulability.
An \lstinline$Error$ location is reachable when a queue overflow occurs ($t_{10}$) or a task in the queue explicitly misses its deadline ($t_9$).
Support for parameters can be added with an array \lstinline$p$ such that \lstinline$p[i][j]$ holds the \lstinline$j$th parameter of \lstinline$q[i]$.

\paragraph{Channels}
The \lstinline{start} channel is for starting execution of a task ($t_3$ and $t_8$), which later gives up the processor with a signal on \lstinline{finish} channel ($t_{11}$, $t_{12}$ and $t_{13}$). 
The \lstinline{invoke} channel is for sending/receiving messages (async call); $t_5$ generates a new task with the deadline given in the variable \lstinline{deadline} by putting the message name, sender and the associated label in the queue.
The \lstinline{delegate} channel generates a sub-task ($t_4$), which inherits the deadline of the parent task.
A reply to an async call is reported on \lstinline{reply} channel ($t_6$).
The blocking statement \lstinline$t?$  passes control to the scheduler using \lstinline{wait} channel ($t_2$); this allows the scheduler to perform a synchronous call when needed. A blocked task is resumed using \lstinline{resume} channel ($t_1$).
To avoid context-switch delays, \lstinline{start} and \lstinline{resume} defined urgent.

\paragraph{Start-up}
An object is initialized by putting `init' and `run' methods in its queue ($t_7$), immediately followed by executing the `init' method ($t_8$).
During `init', the object may receive other messages which are allowed to compete with the `run' method for execution.


\paragraph{Context-switch}
Tasks can have enabling conditions, which may include the availability of a reply, but does not depend on clock values. Therefore, we can define in \Uppaal\ a boolean function \lstinline$isEnabled$ to evaluate the enabling condition for each method when needed.
Whenever a task finishes, the scheduler selects another {\em enabled} task, based on its strategy ($t_{12}$), sets \lstinline$run$ to point to this task, and executes it ($t_3$).
There are two special cases: (1) the last task in the queue has just finished ($t_{13}$); since \lstinline$run$ is zero and \lstinline$q[0]$ gets \lstinline$EMPTY$ after \lstinline$shift$, the processor becomes idle until a new task arrives ($t_5$).
(2) all remaining tasks in the queue are disabled ($t_{11}$); \lstinline$run$ is set to \lstinline$MAX$ to block the processor. In this case, the object may be enabled either by receiving a new task ($t_5$) or receiving a reply signal ($t_6$). The functions \lstinline{insertInvoke} and \lstinline{invokeReply} take care of these cases.

\paragraph{Labels}
The array \lstinline$labels$ associates a boolean to each label. Upon completion of a method, the label used for its invocation is set to true in the \lstinline{shift} function ($t_{11}$, $t_{12}$, $t_{13}$). Since these transitions fire also at processor release points, the function \lstinline{shift} uses the variable \lstinline{complete}. This variable is true by default and will be reset to false at release points.

\paragraph{Synchronous call}
Executing the blocking \lstinline$t?$ statement (\lstinline$wait$ channel) on a remote call blocks the object until a reply is received. However, if \lstinline$t$ is associated to a local call, the scheduler will start the called method (modeling synchronous function call); upon termination of the called method, the scheduler will resume the blocked process (\lstinline$resume$ channel). In order to allow nested synchronous invocations, each message in the queue stores a pointer to the caller method.
As explained in the next section, the blocked method goes back to its initial location allowing recursive synchronous function calls. Note that this recursion is bounded with the queue length.

\paragraph{Scheduling strategy}
The selection strategy is specified as a guard on $t_{12}$. Parts of this guard ensure that we consider only non-empty queue elements (\lstinline$i < tail$) containing an enabled task (by calling \lstinline$isEnabled$) different from the currently running one (\lstinline$run$).
Figure \ref{fig:scheduler} compares the remaining deadline of task \lstinline$i$, obtained by \lstinline$d[ca[i]] - x[ca[i]]$, with other tasks in the queue; task \lstinline$i$ is selected when its deadline is strictly less than that of task \lstinline$m$ for \lstinline$m>=i$ and less than or equal for \lstinline$m<=i$.
By replacing deadlines with a priori defined task priorities, we can model fixed-priority-scheduling.
By fixing \lstinline$run$ to zero, we obtain first-come-first-served strategy.
Nevertheless, the model is not restricted to these strategies.

\subsection{A Formal Encoding from Creol Syntax to Timed Automata}
\label{sec:formal}

We define this semantics in terms of a translation algorithm.
The input to this algorithm is a Creol model consisting of class and interface specifications.
The output is one \Uppaal\ model for each class; this \Uppaal\ model consists of only a set of timed automata templates ($\id{TAT}$), one for each method, and the global declarations ($\id{Dec}$).
The system declarations is not generated automatically, because it depends, among others, on the choice of scheduling strategy (cf. Section \ref {sec::analysis}).
We formally define the translation for a class: $[\![  \id{\bf class} ~ \id{C}~( \id{a^*})~ \id{\bf implements}~ \id{i^*} ~ \id{\bf begin} ~\id{v^*} ~\id{m^*} ~\id{\bf end}  ]\!] = (\id{Dec}, \id{TAT})$, where
{\small
\begin{math}
\id{TAT} = \{ (\id{Beh}(I), C\_I, Args,\id{Local}(I)) ~|~ I \in i^*\} \cup
\{(\trule{\id{\bf op}~ N == S}, \id{C\_N}, \id{Args}, \{\})
| ``\id{\bf op}~ N == S\mbox{''} \in m^*\}.
\end{math}
}
Functions $\id{Beh}(I)$ and $\id{Local}(I)$ return the user defined automata and their local declarations for the interface with the name $I$.
All automata templates have as arguments:
{\small
$
Args =  \{ \verb+const int + arg \verb+;+~|~arg:Type \in a^* \} ~ \cup
 \{ \verb+const int self;+ \}
$
}.
First we define automata templates for methods and later $\id{Dec}$ in this section.

\begin{table}
{\small
\begin{tabular}{@{}p{.19\linewidth}|p{.33\linewidth}|p{.40\linewidth}}
Function & Input & Output \\ \hline
$\trule{.}: M \rightarrow A$ & A method & A timed automaton \\ \hline
$\trule{.}: S \times L \times L \rightarrow T \times L \times I \times 2^{N \times G}$ &   A  Creol statement, two automata locations &
Part of a timed automaton (transitions, locations, invariants), set of enabling conditions \\\hline
$\id{Loc}: L \rightarrow S $&  A location in a method automaton & A Creol statement in the method body \\\hline
$\id{LabelReset}: G \rightarrow 2^U $ & A guard & \Uppaal\ update statements\\\hline
$\id{Abs}: E \rightarrow E$ & Creol expression & \Uppaal\ expression\\ \hline
$\id{labels}:  * \rightarrow L $ & overloaded & label names \\ \hline
$\id{mthds}: * \rightarrow N$ & overloaded & method names  \\
\end{tabular}
}
\caption{Some functions used for translation}
\label{tab:function}
\end{table}


\begin{figure*}[tbp]
\begin{center}

\begin{tabular}{
@{\hspace{-5pt}} m{.59\linewidth} @{~~~~~}
m{.135\linewidth}  m{.015\linewidth}  m{.145\linewidth} m{.001\linewidth}
}

{\small
\begin{math}
\begin{array}{l}
\trule{{\bf with} ~N'~{\bf op} ~N~ ==~ S} = \trule{{\bf op} ~N~ ==~ S} = ( L\cup \{l_0,a,u\} ,l_0,T \cup T_1 , I' )
\mbox{\small ~where}
\\ \trule{S}_{a,u}=(T,L,I,E) \mbox{\small ;~and,~}
I' = \{ (a, c \ge w) ~|~ (a,w) \in I \} \mbox{\small ;~and,~}
\\ T_1 = \{l_0 \xrightarrow[c:=0]{{\it start}[N][{\it self}]?} a, u \xrightarrow{{\it finish}[{\it self}]!} l_0\} \mbox{\small ;~and,}
\\ \mbox{New locations: $l_0$ (initial), $u$ (urgent) and $a$ such that~} \id{Loc}(a) = S
\end{array}
\end{math}
}
{\smallskip}
& &
\includegraphics [scale=\scaleIndex]{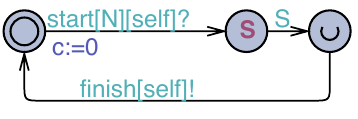}
&&
\\\hline

\noalign{\medskip}

{\small
\begin{math}
\begin{array}{l}
 \trule{S_1;S_2}_{a,e} = \trule{S_1}_{a,l} ~\biguplus~ \trule{S_2}_{l,e} \\
 \mbox{New location: $l$ such that~}\id{Loc}(l) = S_2
\end{array}
\end{math}
}
{\smallskip}
&
\includegraphics[scale=\scaleIndex]{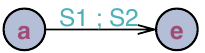} &
$\stackrel{\id{seq~}}{\Longrightarrow}$ &
~\includegraphics[scale=\scaleIndex]{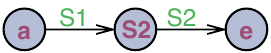} &

\\\hline
\noalign{\smallskip}

{\small
\begin{math}
\begin{array}{l}
\trule{ skip ~\textit{/*@b~@w*/} }_{a,e}  = 
\{a \xrightarrow[c\ge b ~;~ c:=0]{} e\} , \{(a, w)\}
\end{array}
\end{math}
}
&
\includegraphics[scale=\scaleIndex]{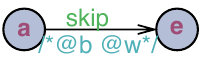} &
$\stackrel{\id{skip}}{\Longrightarrow} $&
~\includegraphics[scale=\scaleIndex]{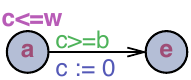} &

{\smallskip}
\\\hline
\noalign{\smallskip}

{\small
\begin{math}
\begin{array}{l}
\trule{ v := ex ~\textit{/*@b~@w*/} }_{a,e}  = 
 \{a \xrightarrow[c\ge b ~;~ c:=0,{\it Abs}(v := ex)]{} e\}, \{(a,w)\}
\end{array}
\end{math}
}
&
\includegraphics[scale=\scaleIndex]{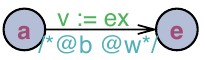} &
$\stackrel{\id{assign}}{\Longrightarrow} $&
~\includegraphics[scale=\scaleIndex]{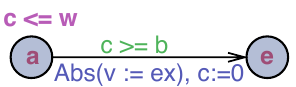} &

{\smallskip}
\\\hline
\noalign{\smallskip}

{\small
\begin{math}
\begin{array}{l}
\trule{ !rec.m()~\textit{/*@b~@w~@d*/} }_{a,e} = 
\{a \xrightarrow[c\ge b~;~c:=0,{\it deadline} := d]{{\it invoke}[0][m][{\it rec}][{\it self}]!} e\},
\{ (a,w)\}
\end{array}
\end{math}
}
&
\includegraphics[scale=\scaleIndex]{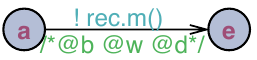}
&
$\stackrel{\id{call~}}{\Longrightarrow} $
 &
\includegraphics[scale=\scaleIndex]{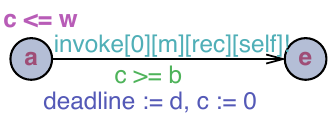}   
&

{\smallskip}
\\\hline
\noalign{\smallskip}

{\small
\begin{math}
\begin{array}{l}
\trule{ t!rec.m()~\textit{/*@b~@w~@d*/} }_{a,e} =
\{a \xrightarrow[c\ge b~;~c:=0,{\it deadline} := d]{{\it invoke}[t][m][{\it rec}][{\it self}]!} e\},
\{ (a, w)\}
\\\mbox {where  $t$ is reused to represent the integer constant for label $t$.}
\end{array}
\end{math}
{\smallskip}
}
&
\includegraphics[scale=\scaleIndex]{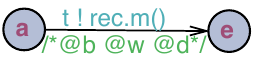}  &
$\stackrel{\id{label}}{\Longrightarrow} $ &
\includegraphics[scale=\scaleIndex]{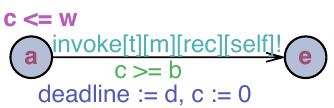} &

\\\hline
\noalign{\medskip}

{\small
\begin{math}
\begin{array}{l}
\trule{ t?~\textit{/*@b*/} }_{a,e}  =
 \{a \xrightarrow[c\ge b]{{\it wait}[t][{\it self}]!} l_0, ~ l_0 \xrightarrow[{\it LabelReset}(t?),c:=0]{{\it resume}[t][{\it self}]?} e\} ,
\{ (a,b)\}
\\\mbox {where  $t$ is reused to represent the integer constant for label $t$.}
\end{array}
\end{math}
{\smallskip}
}
&
\includegraphics[scale=\scaleIndex]{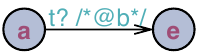}  &
$\stackrel{\id{ret~}}{\Longrightarrow} $ &
\includegraphics[scale=\scaleIndex]{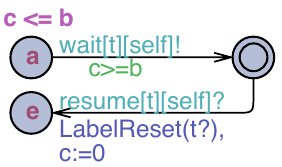}&

\\\hline
\noalign{\medskip}

{\small
\begin{math}
\begin{array}{l}
\trule{ release~\textit{/*@b~@w*/}}_{a,e} = \\
  \{a \xrightarrow[c \ge b ~;~{{\it complete}[{\it self}]:={\it false}}]{delegate[x1][{\it self}]!} u,
~  l_0 \xrightarrow[c:= 0]{start[x1][{\it self}]?} e \}
, \{(a,w)\}, \{(x1,true)\} \\
\mbox{where $u$ is the unique final location and $x1$ is a new name.}
\end{array}
\end{math}
{\smallskip}
}
&
\includegraphics[scale=\scaleIndex]{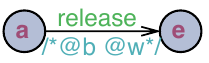}
& $\stackrel{\id{rel~}}{\Longrightarrow} $ &
\includegraphics[scale=\scaleIndex]{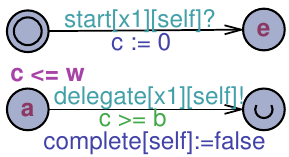} &

\\\hline
\noalign{\medskip}

{\small
\begin{math}
\begin{array}{l}
\trule{ await ~g~\textit{/*@b~@w*/}}_{a,e} = \\
\{ 
~  a \xrightarrow[{c\ge b ~\land~ {\it Abs}(!g) ~;~ {\it complete}[{\it self}]:={\it false}}]{{\it delegate}[x1][{\it self}]!} u,
l_0 \xrightarrow[c:= 0]{start[x1][{\it self}]?} e,
\\\hspace{8pt}
a \xrightarrow[c\ge b ~\land~ {\it Abs}(g) ~;~ c:=0,LabelReset(g)]{~} e 
\} 
,\{(a, w)\},\{(x1,Abs(g))\}\\
\mbox{where $x1$ is a new name, $l_0$ is the initial location and $u$ is the unique final location.}\\
\end{array}
\end{math}
{\smallskip}
}
&
\includegraphics[scale=\scaleIndex]{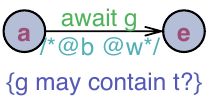}  &
$\stackrel{\id{crel~}}{\Longrightarrow} $ &
\includegraphics[scale=\scaleIndex]{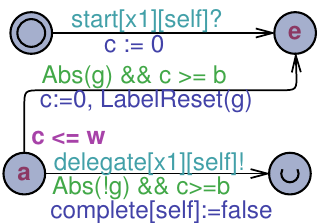} &

\\\hline
\noalign{\medskip}

{\small
\begin{math}
\begin{array}{l}
\trule{ {\it if}~(g)~{\it then}~S~{\it else}~T~~\textit{/*@b~@w*/} }_{a,e} = 
\trule{S}_{l_1,e} ~\biguplus~ \trule{T}_{l_2,e} ~\biguplus~ \\
 ( \{a \xrightarrow[c \ge b ~\land~{\it Abs}(g)~;~c:=0]{~} l_1, a \xrightarrow[c \ge b ~\land~{\it Abs}(!g)~;~c:=0]{~} l_2\} 
,\{l_1,l_2\}, \{(a, w)\}, \{\})\\
\mbox{New locations: $l_1$ and $l_2$ such that }\id{Loc}(l_1) = S;\id{Loc}(e) \mbox{~and~} \id{Loc}(l_2) = T;\id{Loc}(e)
\end{array}
\end{math}
{\smallskip}
}
&
\includegraphics[scale=\scaleIndex]{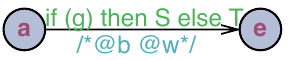} &
~$\stackrel{\id{if~}}{\Longrightarrow} $ &
\includegraphics[scale=\scaleIndex]{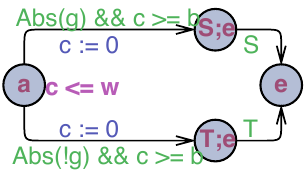} &

\\\hline
\noalign{\medskip}

{\small
\begin{math}
\begin{array}{l}
\trule{ {\it while}~(g)~{\it do}~S~{\it od}~\textit{/*@b~@w*/}}_{a,e} = 
\trule{S}_{l,a} ~\biguplus~ \\
(\{a \xrightarrow[c \ge b ~\land~{\it Abs}(g)~;~c:=0]{~} l, a \xrightarrow[c \ge b ~\land~{\it Abs}(!g)~;~c:=0]{~} e\} 
, \{l\}, \{(a, w)\}, \{\}) \\
\mbox{New location: $l$ such that }\id{Loc}(l) = S; \id{Loc}(e)
\end{array}
\end{math}
}
&
\includegraphics[scale=\scaleIndex]{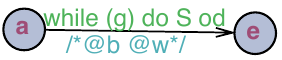}  &
$\stackrel{\id{while}}{\Longrightarrow} $ &
\includegraphics[scale=\scaleIndex]{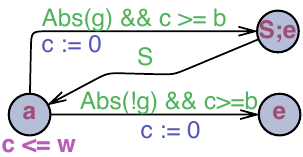} & 

\end{tabular}
\\
\end{center}
\caption{The rules for translating methods to timed automata. 
For each rule, we write only non-empty sets.
The union of two tuples is defined as  $(T, L, I, E) ~ \biguplus ~ (T', L', I', E') = (T \cup T', L \cup L', I \cup I', E \cup E')$
}
\label{fig::generation}
\end{figure*}

\paragraph{Automata Templates}
The timed automata for each method is obtained using the function $\trule{.}$ defined in Figure \ref{fig::generation}.
For the translation of each method, the locations $l_0$ and $u$ refer to the unique locations representing the {\em initial} and {\em final} location of the method automaton, respectively.
The final location $u$ is urgent and is connected to $l_0$ to allow multiple incarnations of the method execution.

The partial function $Loc: L \rightarrow S$ assigns to the locations of a method automaton, their corresponding statements in the method body. This function is not defined for the  initial and the urgent final locations in the generated method automata. With each automaton corresponding to exactly one method, one can use this function to trace from automata back to methods.
In Figure \ref{fig::run}, this function is depicted as labels on locations, referring to line numbers in Figure \ref{fig::creolCode}.

The function $\trule{S}_{a,e} = (T, L, I, E)$ translates the statements $S$ to a set of transitions $T$ on locations $L$ and with location invariants $I$.  Processing $S$ should start from the location $a$ and finish at $e$ (obviously $L$ will contain $a$ and $e$).
Additionally, $E$ will correspond to a set of pairs $(n, en)$, where $n$ is the name of a sub-task (generated by a release point) with the enabling condition $en$ (e.g., see tables in Figure \ref{fig::run}).

Given a method ${\bf op} ~N~ ==~ S$, the corresponding timed automaton is computed as shown in Figure~\ref{fig::generation} by computing $\trule{{\bf op} ~N~ ==~ S}$, which uses the overloaded function $\trule{.}$ to compute the automaton transitions.
A cointerface given using ${\bf with}$ keyword is ignored.
The intuition behind the translation function is given on the right hand side of Figure~\ref{fig::generation} using a notation similar to graph transformation rules.

The translation uses some helper functions.
As explained in the scheduler, there is a global boolean array $labels$.
For every label $t$ that is checked in guard $g$, $LabelReset(g)$ resets $labels[t]$ to false.
$Abs(ex)$ is the abstract of the expression or assignment $ex$ with proper renaming of the variables; the abstract must map the expressions to a finite domain, mapping everything to $true$ is correct but not optimal.
If the domains of the variable are finite we can leave them unchanged, however if they are possibly infinite then we must approximate them.  The user of our translation can decide the exact approximation, as it will need to balance between state-space size and accuracy. A possible definition of these functions are:

{\footnotesize
\begin{tabular}{@{}l@{~~}l@{~~}l}
\\[-6pt]
$LabelReset(g \land g')$& = &$LabelReset(g) \cup LabelReset(g')$\\
$LabelReset(!g) $& = &$ LabelReset(g)$ \\
$LabelReset(l?) $& = &$ \verb+labels[l][self]=false;+$ \\
$LabelReset(b) $& = &$ \{\}$\\[6pt]
\end{tabular}
\begin{tabular}{l@{~~}l@{~~}l}
$Abs(ex) $&=&$ ex \{ \verb+labels[l][self]+ /  \verb+l?+ \} \{ \verb+v[self]+ /  \verb+v+ \}$ \\
&&for variables $v$ and labels $l$.
\end{tabular}
}

Next, we explain two most complex rules with the example in Figure \ref{fig::run}.

\paragraph{ret.}
The statement \lstinline{t?} stops the current method and triggers the scheduler (on \lstinline{wait} channel). If \lstinline$t$ is associated to a local call, the scheduler immediately starts the called method (i.e., sync call). This can be a recursive call because the $ret$ rule moves the method to the initial location $l_0$ (transition from 16 to $l_0$ in  Figure \ref{fig::run}). If \lstinline$t$ refers to a remote call, the object is blocked (no method is executed). The caller is {\em resumed} when the callee has finished and a reply signal is available (transition from $l_0$ to 17 in Figure \ref{fig::run}). Since the upper bound on this statement depends on the called methods, we can only express a lower bound (\lstinline{/*@b*/})  on its execution time.

\paragraph{crel.}
With the \lstinline{await g} statement, if \lstinline$g$ holds, the processor is not released and the method continues (e.g., transition 25 to 26 in  Figure \ref{fig::run}).
If \lstinline$g$ does not hold, control moves to the final location $u$ (e.g., from location 25); therefore, the current task will finish. Instead, it generates a subtask \lstinline$x1$ using the \lstinline$delegate$ channel (thus it inherits the deadline), with enabling condition \lstinline$g$. The transition \lstinline$start[x1][self]?$ defines the subtask \lstinline$x1$ such that it will execute the rest of the original task (e.g., from $l_0$ to 26).

\paragraph{Global Declarations}
In Figure \ref{fig::helper}, we define some helper functions we need for computing the global declarations $Dec$ for `$ \id{\bf class} ~ \id{C}( \id{a^*})~ \id{\bf implements}~ \id{i^*} ~ \id{\bf begin} ~\id{v^*} ~\id{m^*} ~\id{\bf end}$'.
To save space, we do not write the real-time execution information when not relevant for these function definitions.
We use  $E(m)$ as a short hand to return the element $E$ returned by $\trule{m}$ for $m \in m^*$.

%
%

\begin{figure}
\begin{center}

\begin{tabular}{@{~}l@{\hspace{-2cm}}l}
$\id{starts} ~=~ \{E(m) ~|~ m \in m^*\}$\\
$\id{tasks} ~~=~ \{N~|~\id{\bf op} ~N == S \in m^*\} ~ \cup ~ \{n ~|~ (n,en) \in E(m), m \in m^*\}$\\[6pt]

$\id{labels} ~=~ \bigcup \id{labels}(S) \mbox{~ for ~}\id{\bf op} ~N == S \in m^*$ &
$\id{methods} ~=~ \bigcup \id{mthds}(S) \mbox{~ for ~}\id{\bf op} ~N == S \in m^*$\\
{\small
\begin{tabular}{@{} l @{~} c @{~} l}
$\id{labels}(s;S) $&$=$&$ \id{labels}(s) \cup \id{labels}(S) $\\
$\id{labels}( \id{\bf if}\ b\ \id{\bf then}\ S_1\ \id{\bf else}\ S_2 ) $&$=$&$ \id{labels}(S_1) \cup \id{labels}(S_2)$\\
$\id{labels}(\id{\bf while}\ b\ \id{\bf do} ~S~ \id{\bf od} ) $&$=$&$ \id{labels}(S)$\\
$\id{labels}(t!p()~ \id{/*}\id{@d}\id{*/}) $&$=$&$ \{t\}$\\
$\id{labels}(\_) $&$=$&$ \emptyset$
\end{tabular}
}
&
{\small
\begin{tabular}{@{} l @{~} c @{~} l}
$\id{mthds}(s;S) $&$=$&$ \id{mthds}(s) \cup \id{mthds}(S) $\\
$\id{mthds}( \id{\bf if}\ b\ \id{\bf then}\ S_1\ \id{\bf else}\ S_2 ) $&$=$&$ \id{mthds}(S_1) \cup \id{mthds}(S_2)$\\
$\id{mthds}(\id{\bf while}\ b\ \id{\bf do} ~S~ \id{\bf od}) $&$=$&$ \id{mthds}(S)$\\
$\id{mthds}(t!x.m ) $&$=$&$ \{m\}$\\
$\id{mthds}(\_) $&$=$&$ \emptyset$
\end{tabular}
}
\end{tabular}

\end{center}
\caption{Encoding helper functions. As parameter $\_$ represents other possibilities.}\label{fig::helper}
\end{figure}

Given a minimum $d_n$, a maximum $d_x$ and an initial value $d_i$ for deadline,
the global declarations $Dec$ are defined as follows.
The global declarations $Dec$ consists of the following elements. For easier reading, we put them in a list. Treating each item as a set, $Dec$ is formally defined as their union.
The functions $\id{IntT}$, $\id{IntL}$ and $\id{IntM}$ below produce unique integer values, starting from zero ($\id{IntL}$ starts from 1) and incrementing by one every time they are called.

\begin{itemize}
\item global constants. The $\#$ function returns the number of elements in a set.\\
{\footnotesize
$\verb+const int MSG =+ ~\id{Max}(~ \#(\id{methods} \setminus \id{tasks}), \#(\id{tasks})~ )$\verb+;+\\
$\verb+const int nObj = +\#(a^*)+1$\verb+;+\\
$\verb+const int LBL = +\#(labels)$\verb+;+
}

\item  the global clock used by all method automata and the deadline variable. \\
{\footnotesize $ \verb+clock c; meta int[+d_n,d_x\verb+] deadline;+$}

\item 
 a unique number for each task and subtask. \\
{\footnotesize $\{\verb+const int +t\verb+ = +\id{IntT}()\verb+;+~|~t \in \id{tasks}\}$ }

\item 
 a unique number for each label, and a boolean array. \\
{\footnotesize $\{\verb+const int +l\verb+ = +\id{IntL}()\verb+;+~|~l \in \id{labels}\}$ }\\
{\footnotesize \verb$bool labels[LBL+1][nObj];$}

\item 
an array for each variable, either a bool or an int.\\
{\footnotesize $ \{  type~name~\verb+[nObj];+~|~ name:type \in  v^* ~\}$ }

\item a bool to indicate method completion, helping scheduler decide whether to issue a reply signal.\\
{\footnotesize \verb$bool complete[nObj];$}

\item
a unique number for each method called on the objects provided as arguments.\\
{\footnotesize $ \{  \verb+const int +m\verb+ = +\id{IntM}()\verb+;+~|~m \in \id{methods} \setminus \id{tasks} ~\}$ }

\item
the channels used by the scheduler.
{\footnotesize
\begin{verbatim}
chan delegate [MSG+1][nObj];
chan invoke [LBL+1][MSG+1][nObj][nObj];
urgent chan start [MSG+1][nObj];
chan finish[nObj];
chan wait[LBL+1][nObj];
urgent chan resume[LBL+1][nObj];
chan reply[LBL+1][nObj];
\end{verbatim}
}

\item 
code to help the scheduler start the tasks correctly. \\
{\footnotesize
$   \verb+bool isEnabled (int msg, int self) {+$\\
$~~~~ \{ \verb+if (msg == +n\verb+) return +en\verb+;+~~~~~~~|~$ for all $(n,en) \in starts$ \}\\
$\verb+  return true;+$ \\
\verb+}+
}

%


\end{itemize}

\section{Analysis of Real-Time Objects}\label{sec::analysis}

The generated timed automata fit our automata-theoretic framework for modular schedulability analysis of asynchronous objects~\cite{Jaghoori09jlap,JaghouriDBC08RTSS}. The extensions to the original framework  are as follows.
Methods (and their corresponding messages) have enabling conditions.
The completion of each method is reported back to the caller with a reply signal; this enables modeling Creol synchronization mechanisms.
Static labels are used to match replies with their originating calls.
A blocking synchronization statement waiting for reply from a local call leads to a deadlock; instead, as in basic Creol, we transform this situation to the  synchronous execution of the called method.
On the generated timed automata, one can perform normal \Uppaal analyses like reachability; using the $\id{Loc}$ function (see Table \ref{tab:function}), automata locations can be traced back to the original Creol code.
As the original framework is intended, one can perform schedulability analysis to check whether called methods can finish within the required deadlines.

\medskip
\noindent {\bf Schedulability Analysis}
An object  is an instance of a class together with a scheduler automaton.
An object is schedulable, i.e., all tasks finish within their deadlines,
 if and only if the scheduler cannot reach the \lstinline$Error$ location with a queue length of $\lceil d_{max} / b_{min} \rceil$,
where $d_{max}$ is the longest deadline for any method called on any transition of the automata and $b_{min}$ is the shortest termination time of any of the method automata \cite{Jaghoori09jlap}.
However, schedulable objects usually need a much smaller queue length in practice.

We can analyze a closed system of multiple objects, but it may lead to state-space explosion.
We can avoid that by analyzing one object in isolation; the modular analysis is explained in detail in \cite{Jaghoori09jlap}.
In this case, we need to restrict the possible ways in which  the methods of this object could be called.
Therefore, we only consider the incoming method calls specified in its behavioral interfaces.
Receiving a message from another object (i.e., an input action in the behavioral interface) creates a new task (for handling that message) and adds it to the queue.
The behavioral interface does not capture (internal tasks triggered by) self calls.
To analyze the schedulability of an object, one needs to consider both the internal tasks and the tasks triggered by the (behavioral interface, which abstractly models the acceptable) environment.

If a class implements multiple interfaces, we check schedulability with all interfaces together.
Intuitively, that is because such a class should be able to take part in the protocols provided by these behavioral interfaces together. This is also the case in the coordinator example.
We can generate the possible behaviors of an object by making a network of timed automata consisting of its method automata, behavioral interface automaton $B$ and a concrete scheduler automaton.

Once an object is verified to be schedulable with respect to its behavioral interface, it can be used as an off-the-shelf component. To ensure the schedulability of a system composed of individually schedulable objects, we need to make sure their real use is {\em compatible} with their expected use specified in the behavioral interfaces.
For the details of checking compatibility, we refer to our previous work \cite{Jaghoori09jlap,JaghouriDBC08RTSS}.


\paragraph{Example}
To be able to perform  analysis on an object in isolation, we need the behavioral interface specifications; we consider the specification in Figure~\ref {fig::periodicBeh}.
The behavioral interfaces, the methods and a scheduler automaton are put together in \Uppaal.
We verified this object for schedulability with an `earliest deadline first' strategy.
As a result, we found out that there must be at least an inter-arrival of \lstinline$SPEED$=25 time units (\lstinline$SPEED$ in Figure~\ref {fig::periodicBeh}); with this inter-arrival time, the methods need a deadline of at least \lstinline$MD$=21 until synchronization is successful.
Methods \lstinline$run$ and \lstinline$body$ meet their given deadlines of 10 and 50.
In this case, we observe that no more than 7 queue slots is needed.
It is interesting that a jitter for each of the messages  causes nonschedulability, because in the long run, this message will be either too early or too late at some point for synchronization to take place in time.

We can furthermore model check the correctness of the algorithm.
First of all, we can check whether \lstinline{m1} can go through its both release points; this is done by checking the reachability of location marked 26 in \lstinline{m1} automaton.
Furthermore, we added counters for each method, counting the number of times it has synchronized. We can thus check whether, for instance, \lstinline$m1$ can be in its third round, while \lstinline$m2$ has performed only one round. To ensure that \lstinline$m1$ is triggered fast enough, we add three instances of \lstinline$m1$ at time zero.
We observe that such a scenario is indeed not possible.

\section{Conclusions and Future Work}
\label{sec:conc}
We bridge the gap between high-level declarative automata theory and object-orientation; we elevate scheduling that is normally deeply buried in deployment infrastructure up in high-level modeling.
Creol is a full-fledged object oriented modeling language
strong in formal modeling but the great amount of nondeterminism makes model checking impractical.
In this paper, we have extended Creol with real-time and the possibility of specifying scheduling strategies as a complementary feature; nondeterminism is reduced, among others, by specifying scheduling strategies.
We presented a timed-automata semantics, which can be used for verifying real-time properties and in particular schedulability and timed reachability.
This semantics supports features in Creol like processor release points, replies and synchronous method calls, multiple interfaces and loops.


To allow verification we only allow finite state models. We can put a bound on the length of the queues for schedulable systems; therefore using a finite queue length is not an issue.
However, abstracting possibly infinite variables to finite ones may lead to transitions that would not be possible in the original rewrite semantics of Creol. This can only add possible behaviors, with respect to the rewrite semantics, therefore if we say a Creol program meets its deadlines we can be sure that it does.

The automata model of a class consists of the following: one automaton for every method defined in the class, behavioral interfaces describing the overall input/output behavior and a scheduler automaton with the desired scheduling strategy. Using automata, behavioral interfaces can specify non-uniformly recurring tasks rather than for instance periodic tasks or using pessimistic approximations \cite{FersmanYi07acc}.



As further work, we are  looking into extending the original rewrite semantics of Creol with real-time for simulation purposes.
Instead of automatically generating automata from Creol, one may manually create abstract automata models corresponding to the Creol models, like \cite{BoerGJSY09}; to ensure schedulability of the corresponding Creol model, we will study conformance between Creol and timed automata.

\bibliographystyle{eptcs}
\bibliography{Mmajid}

\end{document}